# Bridging Molecular Structure with Macroscopic Electro-Optic Properties *via* Absorption Wavelength and Chromophore Shape

Ksenia M. Malashkeevich,[a,b*] Ilya V. Prolomov[b], Do Khanh Linh[a] and Michael G. Medvedev[b]

[a]Skolkovo Institute of Science and Technology, Bolshoy Boulevard 30, bld. 1, Moscow 121205, Russia

[b]N. D. Zelinsky Institute of Organic Chemistry, Russian Academy of Sciences, Leninsky prosp. 47, Moscow, 119991, Russian Federation

Organic electro-optic (EO) modulators require films in which dipolar chromophores are aligned by an electric field (poling): the better they align in the presence of the field, the higher the EO coefficient $r_{33}$. $r_{33}$ is sensitive to inter-chromophore interactions, so its estimation is challenging and often relies on statistical-mechanical simulations. Here we show that inter-chromophore interactions can be represented by the product of molecule's volume and its sphericity ratio, which we name $Score_{poll}$: the larger this value is, the bulkier the molecule is and the less strongly intermolecular interactions suppress the macroscopic electro-optic response of the poled film. Together with the maximum EO coefficient ($r_{max}$), which can be estimated directly from maximum absorption wavelength, $Score_{poll}$ can be used to estimate $r_{33}$ measured under comparable poling conditions.

Electro-optic (EO) modulators convert electrical signals into optical ones and are key elements of optical data links (Fig. 1). Many high-speed modulators rely on the linear EO (Pockels) effect, in which an applied field $E$ changes the refractive index $n$ of the active material by $\Delta n = -½n^3r_{33}E$, where $r_{33}$ is the EO coefficient. A larger $r_{33}$ lowers the drive voltage and makes the device more miniature.[1] Lithium niobate, the industry standard, has $r_{33} \approx 30$ pm V$^{-1}$. Films of organic donor–π–acceptor (D–π–A) chromophores now reach $r_{33}$ above 1000 pm V$^{-1}$ and have been integrated into silicon–organic and plasmonic–organic hybrid devices.[1–3]

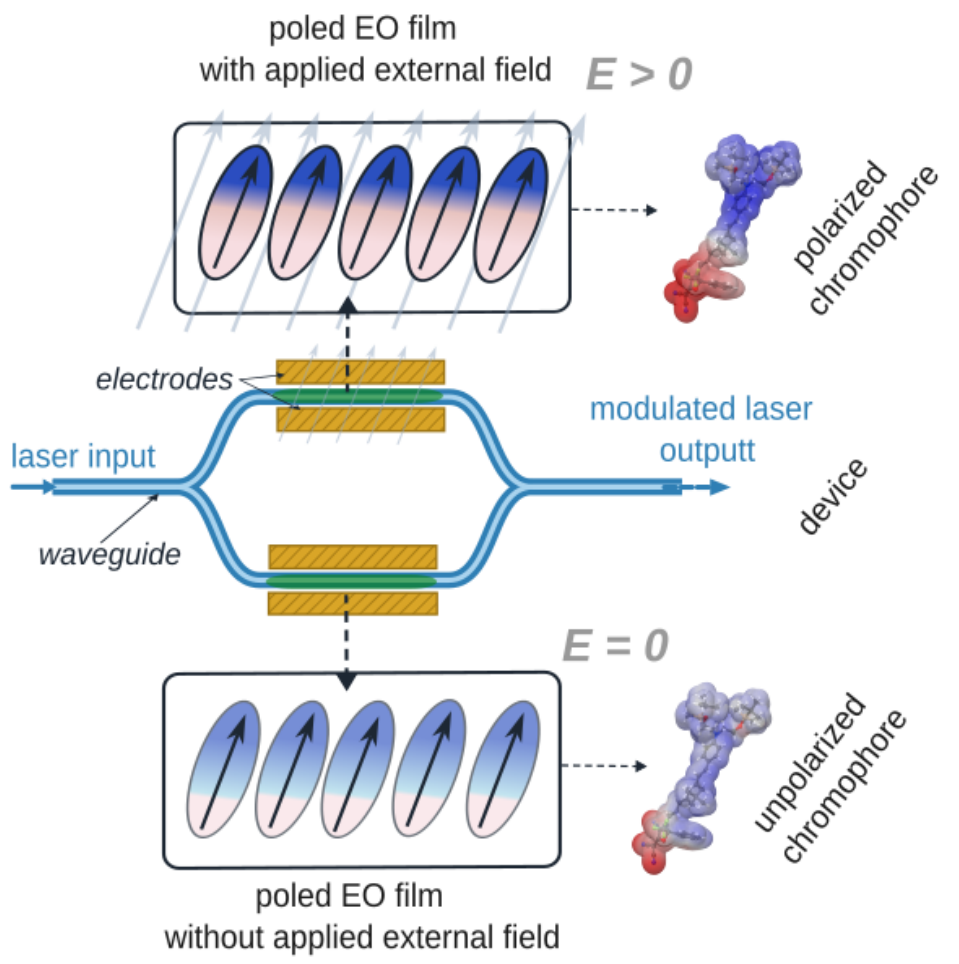

**Fig. 1 From molecule to device: electro-optic modulation across length scales.** Middle: a Mach–Zehnder electro-optic modulator. Top and bottom: a poled chromophore ensemble with and without an applied modulation field (E > 0 and E = 0), respectively; ellipsoids represent chromophores and arrows indicate their dipole moments. Molecular views show the electrostatic potentials of YLD124 calculated with and without a uniform field of 0.002 a.u. ($1.0\times10^9$ V m$^{-1}$) along the donor–acceptor axis. CHELPG-derived electrostatic potentials were calculated at the CAM-B3LYP/def2-TZVPD level with CPCM solvation (chloroform) and mapped onto the 0.002 e bohr$^{-3}$ electron-density isosurface (red, negative; blue, positive).

Such films show the Pockels effect only after poling which aligns their dipolar chromophores: the film is heated close to its glass transition temperature, a strong direct current (DC) field orients the chromophore dipoles and cooling with the field still applied freezes their orientation.[4]
The $r_{33}$ of a polymer film containing organic chromophores can be estimated using Eq. 1.[4–6] There $\rho_N$ is the number density of chromophores, $\beta(-\omega, 0, \omega)$ is their hyperpolarizability for the Pockels effect, $n$ is the refractive index, $g(\omega, \varepsilon)$ is a local-field factor and $\theta$ is the angle between the final chromophore dipoles and the poling field. The acentric order parameter $\langle\cos^3\theta\rangle$ is zero for an isotropic film (before poling) and is one for perfect polar order; real poled films usually have $\langle\cos^3\theta\rangle$ of ~0.1-0.25[6]. It depends on the poling conditions, the host, the chromophore loading and, above all, the electrostatic interactions that might induce antiparallel pairing of the chromophore molecules.[7,8] This collective property is beyond single-molecule calculations; even condensed-phase simulations give highly uncertain absolute values.[6,8]

$$r_{33} = \frac{2g(\omega,\varepsilon)}{n^4}\rho_N \beta(-\omega,0,\omega)\langle\cos^3\theta\rangle \quad (1)$$

Enhancing the molecular first hyperpolarizability, β, is the central objective in the computational design of organic EO chromophores.[2,9] However, a large β does not necessarily result in a large $r_{33}$ (Eq. 1), since it usually comes with a large ground-state dipole moment $\mu$, which induces strong dipole–dipole interactions between chromophores resulting in their anti-parallel aggregation. This aggregation reduces $\langle\cos^3\theta\rangle$ in Eq. 1, negating the expected $r_{33}$ improvement.[4,7,8] The most efficient chromophores therefore combine a strong D–π–A core with bulky groups that keep neighbouring dipoles apart.[2,3,10]
While $r_{33}$ estimation requires very expensive calculations including molecular dynamics, its upper limit $r_{max}$ can be estimated directly from molecule's maximum absorption wavelength $\lambda_{max}$, as derived by Tillack (Eq. 2)[6]. He considered all chromophores to be aligned exactly ($\langle\cos^3\theta\rangle = 1$), took $2g/n^4$ to equal 1, used Kuzyk's limit[11] for the off-resonant hyperpolarizability ($\beta_{max} \propto N^{3/2}\lambda_{max}^{7/2}$, where $N$ is the number of π-electrons), and applied the largest number density compatible with the size of a π-system, $\rho_{N,max} \propto (N\lambda_{max})^{-3/2}$:

$$r_{max} = \frac{2^{3/2}3^{1/4}e^3}{\varepsilon_0(hc)^2}\lambda_{max}^2 \approx 43.8\times10^{-3}\lambda_{max}^2 \text{ pm V}^{-1} \ (\lambda_{max} \text{ in nm}) \quad (2)$$

The dependence on the number of π-electrons cancel out, so $r_{max}$ depends only on $\lambda_{max}$. However, $r_{33}$ of real chromophores usually reach < 2% of $r_{max}$ (JRD1, YLD124, and related chromophores)[6] with large deviations due to the chromophore aggregation, which Tillack's approximation does not take into account.

In this article we propose $\text{Score}_{poll}$ as a simple measure of chromophore's tendency to aggregation computed directly from molecular structure without the need for molecular dynamics. We find that product of $\text{Score}_{poll}$ and $r_{max}$ correlates well with experimental $r_{33}$ values of 5 distinct chromophores measured under similar conditions, while $r_{max}$ alone and β exhibit worse correlations.

We define $\text{Score}_{poll}$ as shown in Eq. 3. There $V_{cav}$ is the molecular cavity volume in a continuum solvent model (PCM) and $I_{min}$ and $I_{max}$ are the smallest and largest unweighted normalized principal moments of inertia, respectively. The latter ratio is the larger the less spherical a molecule is, and so is usually called sphericity ratio[12]. Molecular cavity volume was taken from the ORCA output, and spericity ratio was computed directly from the optimized molecular structures (see below).

$$\text{Score}_{poll} = V_{cav}\frac{I_{min}}{I_{max}} \quad (3)$$

To test usefulness of $\text{Score}_{poll}$ we have selected five chromophores with a tricyanofuran acceptor that were designed for a large $r_{33}$ and measured under similar conditions: JRD1, YLD124, BAY1, BAY9, and BAF1 (Fig. 2).[2,3,10,13] Because $r_{33}$ depends strongly on the poling conditions, we used, wherever possible, values measured at 1310 nm for neat films poled with a $TiO_2$ charge-blocking layer. Chromophores BAY10 and BTF1, also shown in Fig. 2, were also synthesized by the same group, but their $r_{33}$ was not experimentally measured under similar conditions. All seven chromophores share the acceptor (red in Fig. 2) but differ in the donor (blue), the π-bridge (green), and therefore in both conjugation length and molecular shape.

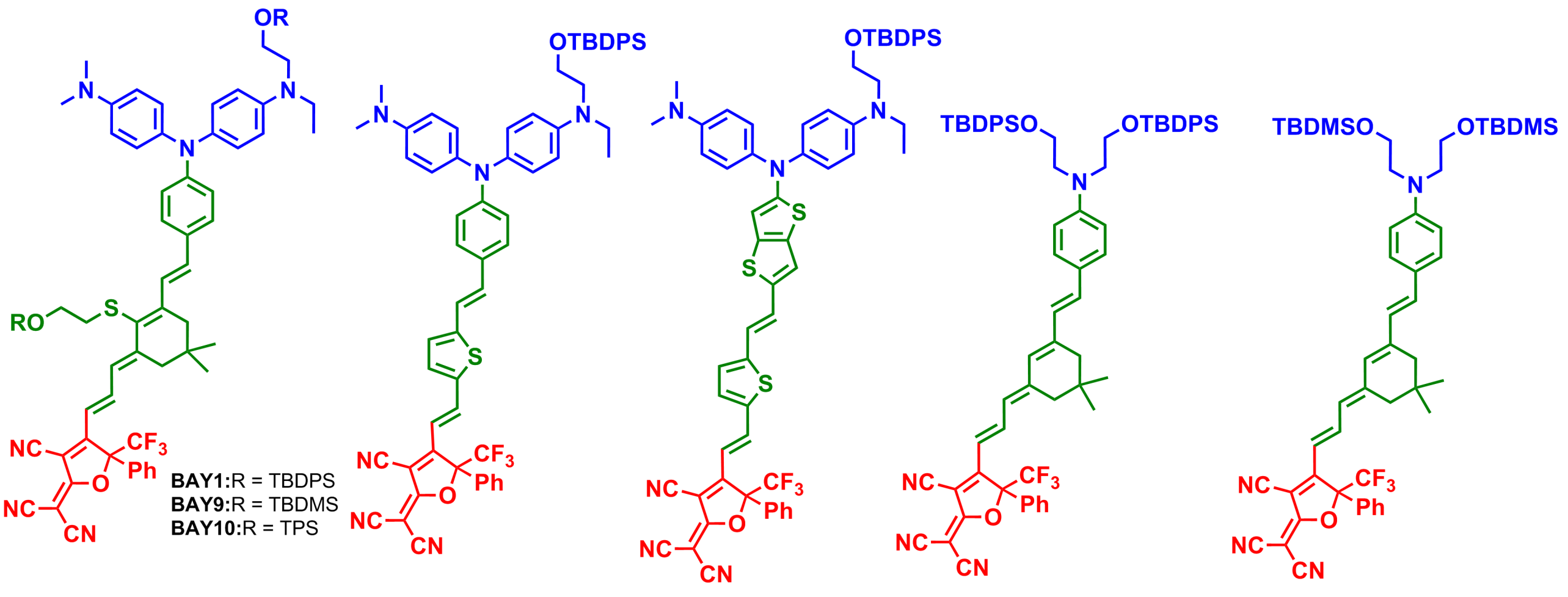

**BAY family** **BAF1** **BTF1** **JRD1** **YLD124**

**Fig. 2** Molecular building blocks and composition of the studied tricyanofuran chromophores - BAY1, BAY9, BAY10, BAF1, BTF1, JRD1, YLD124. Chemical structures incorporate the donor (blue), π-conjugated bridge (green) and acceptor fragments (red). TBDPS, *tert*-butyldiphenylsilyl; TBDMS, *tert*-butyldimethylsilyl; TPS, triphenylsilyl.

The complete computational pipeline used in this work is shown in Fig. 3. For each molecule we first performed a conformational search using the global optimizer GOAT[14] with the semiempirical GFN2-xTB method[15] in implicit chloroform. It generated about 2000 structures per molecule, depending on molecular flexibility. The CREGEN routine of CREST[16] kept structures within 3 kcal $mol^{-1}$ of the global minimum and selected 10 with the most distinct π-bridge geometries. They were re-optimized at the $r^2$SCAN-3c[17] level of theory in the conductor-like polarizable continuum model (CPCM)[18,19] of chloroform, which left 2-8 conformers within 3-4 kcal $mol^{-1}$. CPCM cavity volume $V_{cav}$ and the inertia tensor required for $Score_{poll}$ were computed at the same level of theory. For each selected conformer we computed $\lambda_{max}$ with TD-DFT at the CAM-B3LYP/def2-TZVPD/CPCM level of theory[20,21]. Static hyperpolarizability tensor ($\beta$) values were computed for the same selected conformers using M06-2X[22] density functional, which was used for their computation by Xu et al.[2], Kalinin et al.[23], and Levitskaya et al.[24]. All calculations were performed in ORCA 6.1.1.[25,26] Cavity volumes, $\lambda_{max}$, and the scalar shape parameter $A = (I_{max}/I_{min})^{1/2}$ were averaged using Boltzmann weights derived from $r^2$SCAN-3c electronic energies at 298.15 K. The ensemble descriptor was defined as $Score_{poll} = \langle Vcav\rangle/\langle A\rangle^2$. See ESI for full computational details including unexpected sensitivity of $\beta$ to the conformer, which hinders direct calculation of $r_{33}$ even when data on $\langle \cos^3\theta \rangle$ is available.
While values of $r_{max}$ and β themselves correlate decently with $r_{33}$ ($R^2$ of 0.72 and 0.62, respectively, Figure 3, left), they measure the same polarization ability of the molecule and do not address its aggregation ability, so they place some chromophores with drastically different $r_{33}$'s (e.g., BAF1 and BAY1) very close.

$Score_{poll}$ as we defined it rewards thick bulky molecules and penalizes thin needle- and disk-shaped ones (Fig. 3). Molecules of a thin chromophore can lie side by side and form a tightly bound antiparallel pair with almost no net dipole, which the poling field cannot orient. In a thick molecule, on the other hand, bulky groups surround the π-bridge and keep neighboring dipoles apart; the cavity volume in the numerator rewards this site isolation, which underlies design of the BAY family of chromophores.[2] For example, slim YLD124 with compact silyloxy groups only at the donor has $Score_{poll} = 0.66$, while BAY1 tert-butyldiphenylsilyl-containing side groups at the donor and the bridge has $Score_{poll} = 1.59$.
Since $r_{max}$ and $Score_{poll}$ describe different effects which are both necessary for a chromophore to exhibit high $r_{33}$, but differ in units of measurement, we combined them by multiplication. Strikingly, the resulting product of $r_{max}$ and $Score_{poll}$ has shown a very good correlation with $r_{33}$, with the $R^2 = 0.89$ for five chromophores (Fig.3). To make this product more convenient, we normalized it by its value for JRD1. The equation for predicting $r_{33}$ for a chromophore from the normalized product (P) reads as follows:

$$r_{33} = 664.48 \times P - 289.93 \text{ pm V}^{-1} \quad (4)$$

Introduction of $Score_{poll}$ corrects the positioning of BAF1 relative to BAY1 (see their structures and tensors in Figure 3, top right). The correlation is encouraging, although the small sample size and differences in poling conditions limit its predictive interpretation.

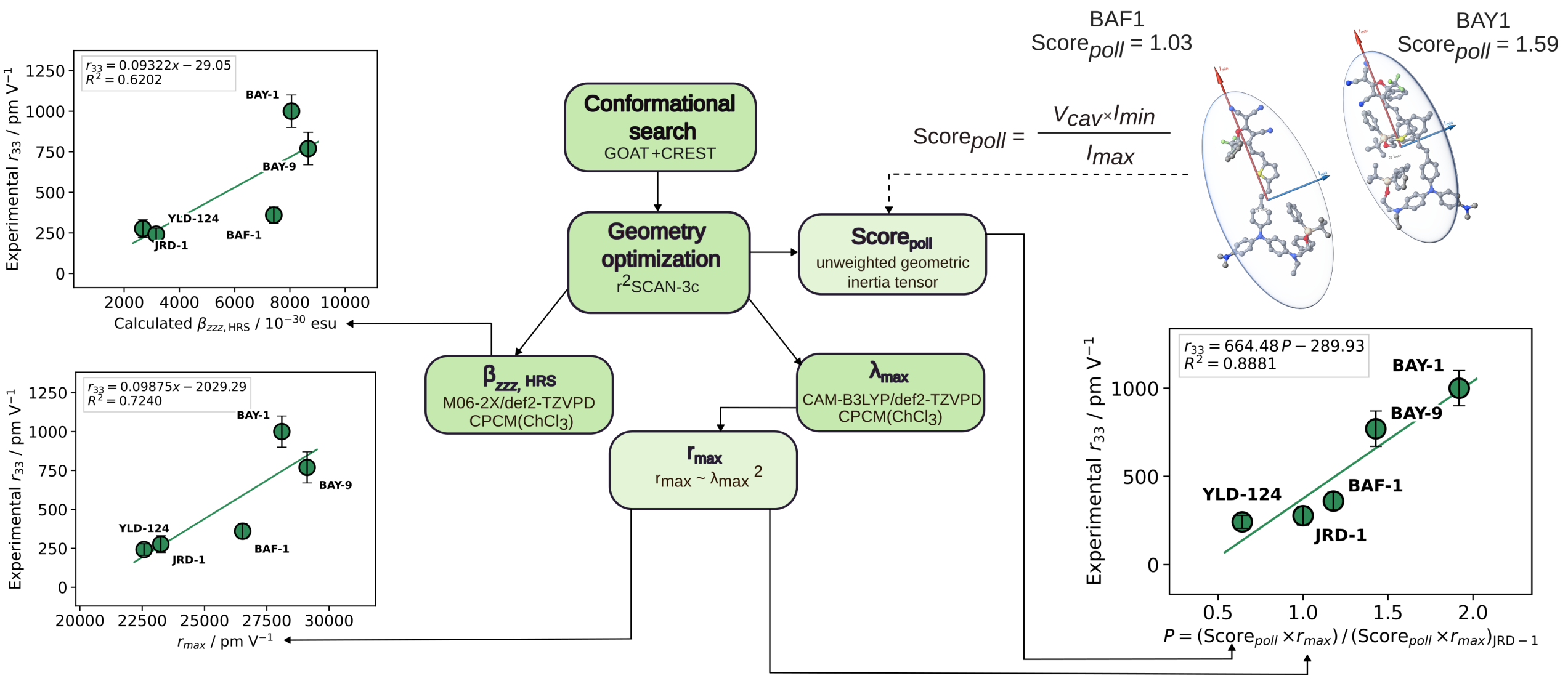

**Fig. 3. Computational workflow and comparison of molecular descriptors with experimental electro-optic coefficients.** Centre: conformational sampling and quantum-chemical calculations used to obtain $\beta_{zzz,hrs}$, $\lambda_{max}$, $r_{max}$ and $Score_{poll}$, with the latter derived from the unweighted geometric inertia tensor. Left: experimental $r_{33}$ versus $r_{max}$ and calculated hyperpolarizability. Bottom right: $r_{33}$ versus the product P = $(Score_{poll} r_{max})/(Score_{poll} r_{max})_{JRD-1}$. Lines show linear least-squares fits to five experimental points; error bars indicate reported experimental uncertainties. Top right: representative conformers of BAF1 and BAY1 illustrating contrasting molecular shapes; the displayed $Score_{poll}$ values are normalized to JRD1.

The obtained correlation also suggests how the two chromophores BAY10 and BTF1 without proper experimental $r_{33}$ data might perform. BAY10, a candidate for experimental testing under matched conditions, has the highest normalized product of $r_{max}$ and $Score_{poll}$ (P = 2.16), suggesting that it may outperform even BAY1, which has both highest P and $r_{33}$ among the considered chromophores with experimental values. BTF1 has the highest $r_{max}$ (1.31 relative to JRD1), consistent with the extended conjugation provided by its thienothiophene unit; however, its low $Score_{poll}$ (0.79) limits the combined descriptor to P = 1.04. This suggests that adding bulky groups around its bridge to hinder antiparallel pairing may be more beneficial than further red-shifting its absorption.

Three assumptions limit the scope of our model. First, $Score_{poll}$ depends only on the shape of the molecule, without taking into account the host, loading and poling conditions, all of which affect $\langle \cos^3\theta \rangle$.[7,27] Thus, the product P can be expected to correlate well with experimental $r_{33}$ only for films prepared under similar conditions. Second, $r_{max}$ accounts only for off-resonant hyperpolarizability, so it ignores both the resonant enhancement of $r_{33}$ and the optical loss of chromophores that absorb close to the operating wavelength. Thus, the correlation might break for strongly red-shifted candidates. Third, our model assumes that the conformations frozen in the poled film match the ensemble computed in implicit chloroform. While this might be expected to work in general, this assumption may be less transferable when specific host–chromophore interactions alter conformer populations or packing.

## Conclusions

We show that single-molecule calculations can predict the EO coefficient of a poled film. The product of upper limit $r_{max}$ and of the shape-based aggregation resistance $Score_{poll}$ correlates well with experimental $r_{33}$, obtained under comparable poling conditions, for five tricyanofuran chromophores. Our model completely supports the empirical design rule: shift $\lambda_{max}$ to the red end of the spectrum and make the molecule thicker with bulky groups around the bridge. Now this rule has been quantified.

## Author contributions

KMM - methodology, investigation, software, writing - original draft, review & editing, MGM - conceptualization, resources, writing - review & editing, IVP – conceptualization, DHL - investigation.

## Conflicts of interest

There are no conflicts to declare.

## Data availability

The data supporting this article have been included as part of the ESI: conformer ensembles, per-conformer $\beta$ and $\lambda_{max}$, cavity volumes and inertia tensors.

## Acknowledgements

Authors are grateful to the Russian Science Foundation for financial support (grant #22-73-10124-P). This work has been carried out using computing resources of the federal collective usage center Complex for Simulation and Data Processing for Mega-science Facilities at NRC "Kurchatov Institute", http://ckp.nrcki.ru/.

## Notes and references

Electronic supplementary information (ESI) available: computational details, conformer ensembles and inertia tensors.